# Multiscale approaches to high efficiency photovoltaics

**J.P. Connolly[1a], Lejo J. Koduvelikulathu[2], D. Mencaraglia[3], Julio C. Rimada[4], Ahmed Nejim[5], G. Sanchez[1]**

[1] Universidad Politécnica de Valencia, NTC, B 8F, 2º. Camino de Vera s/n. 46022, Valencia, Spain.
[2] International Solar Energy Research Center Konstaz; Rudolf-Diesel str. 15, Konstanz, Germany
[3] Laboratoire de Génie Électrique de Paris, LGEP, UMR 8507 CNRS-Supélec, UPMC, UPS, 11 rue Joliot-Curie, Plateau de Moulon, 91192 Gif-sur-Yvette Cedex, France
[4] Solar Cell Laboratory, Institute of Materials Science and Technology (IMRE), University of Havana, Havana, Cuba
[5] Silvaco Technology Centre, Compass Point, St. Ives, Cambridgeshire PE27 5JL, UK.

**Abstract.** While renewable energies are achieving parity around the globe, efforts to reach higher solar cell efficiencies becomes ever more difficult as they approach the limiting efficiency. The so-called third generation concepts attempt to break this limit through a combination of novel physical processes and new materials and concepts in organic and inorganic systems. Some examples of semi-empirical modelling in the field are reviewed, in particular for multispectral solar cells on silicon (french ANR project MULTISOLSI). Their achievements are outlined, and the limits of these approaches shown. This introduces the main topic of this contribution, which is the use of multiscale experimental and theoretical techniques to go beyond the semi-empirical understanding of these systems. This approach has already led to great advances at modelling which have led to modelling software which is widely known. Yet a survey of the topic reveals a fragmentation of efforts across disciplines, firstly, such as organic and inorganic fields, but also between the high efficiency concepts such as hot carrier cells and intermediate band concepts. We show how this obstacle to the resolution of practical research obstacles may be lifted by inter-disciplinary cooperation across length scales, and across experimental and theoretical fields, and finally across materials systems. We present a European COST Action "MultiscaleSolar" kicking off in early 2015 which brings together experimental and theoretical partners in order to develop multiscale research in organic and inorganic materials. The goal of this defragmentation and interdisciplinary collaboration is to develop understanding across length scales which will enable the full potential of third generation concepts to be evaluated in practise, for societal and industrial applications.

---

[a] Address for correspondence: connolly@ntc.upv.es

# 1. Introduction

Efficiencies in first and second generation cells are somewhat saturated, and have been increasing only slowly for the last few decades [1]. These small incremental increases are nevertheless well worthwhile, because the resulting cost reduction is significant.

Given that even these small efficiency increases are worthwhile, there is all the more impetus to identify new mechanisms capable of bridging the vast gap between obtainable systems efficiensies and the fundamental efficiency limits which remain well above achieved efficiencies.

As a result, great efforts are underway to quantify the achievable efficiency improvements from a number of so-called "third generation" concepts in photovoltaics. The first and best understood is the multijunction or tandem concept. This continues to yield improvements, with the current record standing at 46% for concentrator cells, as reported in late 2015 [1]. For a number of reasons, however, ranging from cost to systems complexity, this has yet to make a major impact for terrestrial photovoltaics.

Following this early high efficiency design, a number of novel concepts have emerged since the nanostructured semiconductor revolution starting in the 1980ies. These high efficiency concepts have in common the use of nanostructures allowing the manipulation of optical, vibrational, and electronic physical properties of the materials constituting the solar cell. The phenomena dictate the length scales which are typically of the order of microns for optical properties, and nanometres for electronic and vibrational properties, while the device scale brings with it a length scale of the order of millimetres.

The most actively studied of these novel structures are the quantum well solar cell (QWSC) which emerged in the 1980ies, the intermediate band solar cell in the 1990ies, and the hot carrier solar cell proposed much earlier, but which led to conceivable physical structures on the 1990ies,

This range of scales requires a multiscale approach in both modelling and experiment for proper study of novel properties of these systems. The multiscale approach allows correct consideration of both microscopic origins of nanostructured properties and resulting impacts on device characteristics. This cross-cutting requirement imposes the need for exchange between theory and experiment, the scope of which surpasses research proposals which typically focus on one scale.

The materials focus has evolved from the traditional efficient and flexible III-V family, to include silicon photonics. In addition, organic semiconductors present an increasingly important challenge due to advances in organic electronics and solar cells.

This paper presents examples of studies of nanostructured cells by semi-empirical and numerical approaches illustrating the need for multiscale approaches. These concern nanostructures solar cells, the first exploring novel physical concepts using semi-empirical analysis and which are difficult to extend to a multiscale approach. The second using numerical techniques which are more flexible and open to multiscale approaches.

# 2. Semi-empirical modelling

A semi-empirical model is one which relies in part on parameterisations of physical properties as an input to mainly analytical solution methods. This restricts the exactness of the physical descriptions that can be applied, but yields greater understanding in those cases where this restricted analytical description applies. This generally allow a better phenomenological understanding of the system being studied, but at a cost in accuracy, although the accuracy may be refined by refining the analytical description.

Such a model has been applied over the years to the quantum well solar cell (QWSC) [2] proposed as one of the earliest third-generation concepts. This is a system where the effective bandgap of the solar cell is lowered in order to absorb more light, while in principle avoiding the loss in open-circuit voltage due to increased recombination which is a natural corollary of lowering the gap and increasing absorption.

Figure 1 schematically shows a *p-i-n* version of the structure showing the main feature which is the undoped quantum well system sandwiched between doped layers such as to maintain a field across the wells at the operating voltage.

The modelling of these systems has been reported elsewhere [3]. In brief, it proceeds by analytical solution of the usual set of equations (Poisson, transport, current continuity) and of the Schrödinger equation in the finite square well approximation for thick barriers, that is without superlattice effects. The modelling reported in [3] uses the resulting analytical carrier profiles in the charge neutral and depletion regions to calculate the recombination rates. The strength of this approach is shown in this work by detailed modelling of the light and dark current characteristics of both bulk and quantum well solar cells, as well as published record triple junction solar cells.

This structure has been generalised to strained materials, where a technique of stress balancing was developed [4]. This allows the growth of successive thin layers with periodic compressive and tensile stress, such that the overall stress can remain nil in principle. In practise, although a slight build up of strain energy remains, many quantum well / barrier periods can be grown using strained materials which would otherwise lead to strain relaxation and generation of defects on a level prohibitive for optoelectronic applications. This technique has therefore permitted the growth of a wide range of QWSCs with strained materials.

This modelling approach relies on published data for most cell materials and transport properties relevant to the solution, hence the name "semi-empirical". The only exception is the non-radiative Shockley-Rea-Hall recombination lifetime, assumed equal for electrons and holes, which is the only fitting parameter.

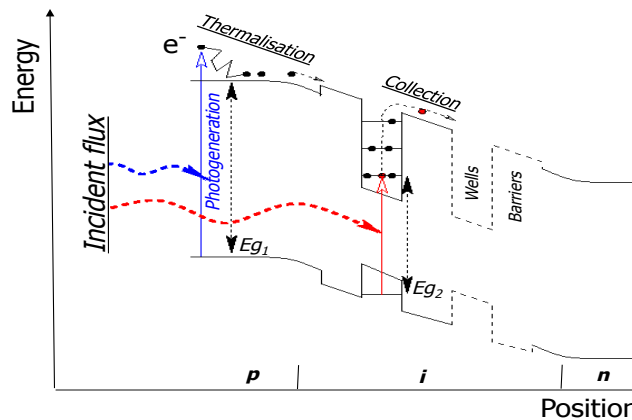

**Figure 1.** Schematic representation of a p-i-n quantum well solar cell (QWSC) showing generation in well and barriers regions, thermalisation, collection, and recombination mechanisms.

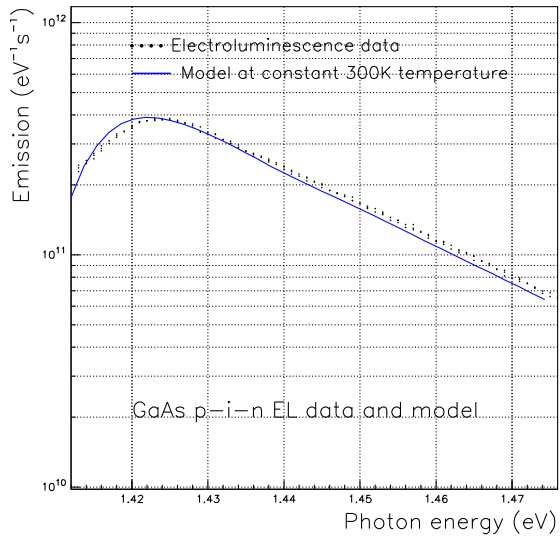 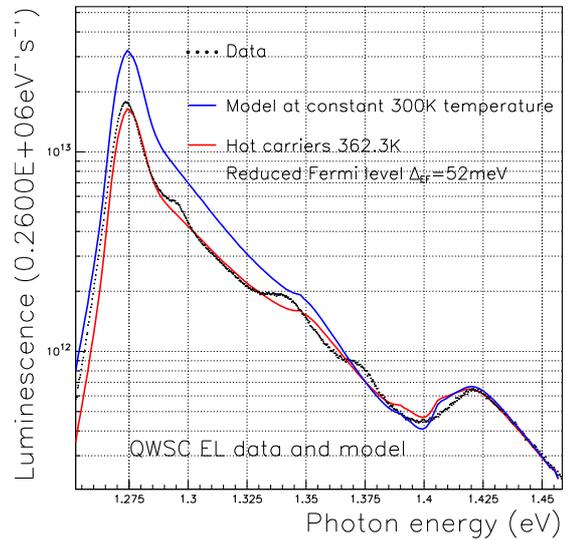

**Figure 2.** Experimental (dots) and theoretical (line) luminescence of a bulk *pin* solar cell showing a good fit with the experimental applied bias for standard temperature of 296K.

**Figure 3.** Experimental (dots) and theoretical fits of a QWSC showing a good fit at room temperature in the bulk (dark line) but a higher temperature needed for well luminescence (red line).

This approach has, in this case, explicitly demonstrated the dominance of radiative recombination in quantum structures, whilst showing that equivalent bulk samples are dominated by non-radiative recombination, providing the first fundamental advantage of the QWSC for practical applications, and in particular showing the advantages of this structure for concentrating photovoltaics.

Of greatest interest from the multiscale perspective is the use of analytical potentials and quasi-Fermi levels separations to evaluate the luminescence from first principles. The luminescence is of particular interest in the QWSC structure given the demonstrated radiative dominance at high bias [5]. This has invited applications in photon recycling in particular [6]. This model of the luminescence is expressed by the Planck grey-body expression for electrons and holes, the populations of which are characterised separate quasi-Fermi levels, with a quasi-Fermi level determined by the applied bias. The temperature of the populations, is in principle be in equilibrium with the lattice given the very rapid thermalisation mechanisms ensuring very rapid thermalisation of carriers compared to recombination lifetimes. Nevertheless, experimental analysis of the luminescence of bulk samples compared to QWSC samples has indicated that the quantum well luminescence shows signs of a greater temperature than the bulk material, details of which are available in [7].

Figure 2 shows the modelled luminescence of a bulk sample showing that the assumption of a quasi-Fermi level separation determined by the applied bias, and a uniform temperature equal to the ambient provides an excellent fit, in particular with the exponential decay with increasing energy being well reproduced.

Figure 3 on the other hand shows a similar good fit for bulk luminescence, but finds that the well luminescence implies a greater temperature and a greater quasi-Fermi level separation, suggesting that the carrier populations in the wells are suppressed.

These intriguing results raise many questions which the semi analytical modelling cannot answer. The roots of this inability to rely on these results stems directly from the absence of any phonon dynamics in the modelling influencing carrier populations in the barrier and well regions of the QWSC. Whilst this assumption is valid for the bulk sample, the same is clearly not true of the QWSC.

The model however lacks the ability to identify the mechanisms which may be responsible for a suppressed carrier concentration and higher carrier temperature in the wells than in the bulk material. As such, it cannot be ruled out that the apparent carrier temperature difference is due to some underlying flaw in the semi-empirical approach. A first suspect is the application of the Planck grey-body formalism to two populations which are assumed to be thermally decoupled.

This brief sketch of intriguing observations in the field of nanostructured solar cells and their analysis in terms of semi-empirical models is therefore a first example of the need for multiscale approaches. In this case, the need is to move on from semi-empirical parameterisations of material

## 3. Numerical modelling

Numerical modelling complements analytical techniques in that, in principle at least, no approximations need to be taken. In practise, limits on numerical methods limit this to some extent, as do the necessarily approximate description of systems in the first place, which are then translated into equations which can be discretised and solved numerically. The advantage of numerical solutions is therefore a more exact and more flexible solution. The disadvantage, conversely, is generally a less physical interpretation of the system being studied. Given the significant investment in time spent developing efficient solution methods, the use of numerical methods is often more efficient if pre-written numerical libraries are used. A much used example is the LAPACK Linear Algebra Package. Furthermore, a number of commercial software solutions exist including SILVACO, COMSOL, and on the multiscale front TIBERCAD.

We will very briefly sketch two examples of numerical modelling in order to raise two examples of short-comings of numerical modelling which may be addressed by multiscale analysis. In both cases the numerical modelling is done in the SILVACO TCAD framework. This paper being concerned with the justification of multiscale modelling, we do not go in to details of the modelling of these structures. It suffices to say that the physical models available in the Silvaco simulator are capable of generality sufficient to deal with this structure in two or three dimensions.

*3.1. MultiSolSi : GaAs on Si multijunction modelling*

The first example of numerical modelling in the context of the MultiSolSi French ANR project. This proposes a novel growth mechanism [8] to grow defect-free GaAs nanocrystals on silicon substrates for multi-junction applications for which preliminary modelling relied on analytical models in order to identify the potential of this structure for multijunction solar cells [9].

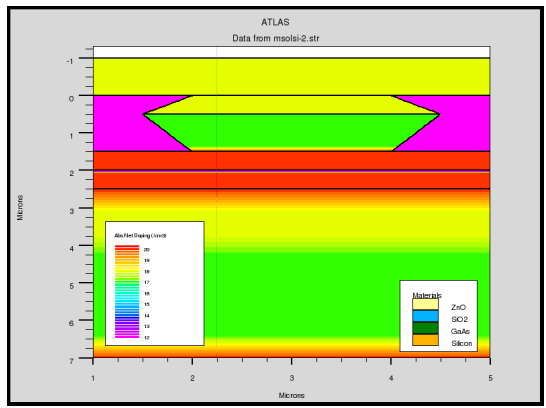 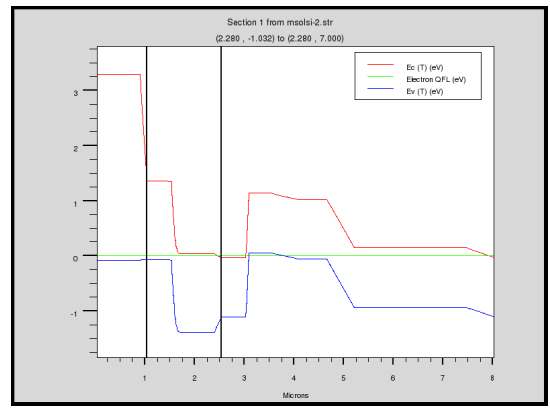

**Figure 4.** Silvaco description of the MultiSolSi schematic tandem solar cell structure showing net doping densities.

**Figure 5.** Calculated band structure of the MultiSolSi structure showing the high bandgap window on the left, and succeeding GaAs and Si solar cells connected in series by a tunnel junction.

The analytical modelling this structure is subject to major approximations, the most important of which are idealised analytical models of tunnel junctions, and the assumption of planar structures suitable for a one dimensional treatment. While these models are in principle capable of including strain effects on opto-electronic parameters, as has been demonstrated for QWSCs, this approach is too limited given the complexity of the problem.

A numerical approach using Silvaco's ATLAS and LUMINOUS simulators is capable of addressing this problem. The definition of the tandem multijunction structure is shown schematically in figure 4, showing only the geometry of the GaAs nano-crystal giving partial coverage of the Si solar cell. The nanocrystal is encased in a transparent matrix and topped by a transparent conducting window layer. Layer thicknesses in this schematic are far from optimal for illustrative purposes. But the doping levels are sufficient to allow a suitable tunnelling current to enable the structure to function as a tandem solar cell. Figure 5 shows the corresponding band-structure for a cross-section through the GaAs micron-scaled crystal where the tunnel junction is clearly vsiible at the interface between the materials.

The first lacking element in this description however is the description of strain modification of material and transport properties, which are absent. This is relevant in the GaAs-Si heterojunction described in detail in the references, where the GaAs is highly strained, although the growth technique employed avoids the stress exceeding the Matthews-Blakesleee strain relaxation limit in the volume, thereby avoiding the formation of strain-induced defects.

In this case the SILVACO model can address such modifications of material properties because arbitrary modification of materials parameters are possible with a sufficiently detailed granular definition of the structure. This is feasible because of the use of numerical solution techniques, within, of course, the limitations of computational power available. Furthermore, it must be noted that in this particular case, this strained region is very thin, and therefore has little effect on the resulting opto-electronic performance of this structure, this is fortuitous.

Despite the work-arounds that can be used to deal with this question it must be noted that the work-arounds are equivalent to the semi-empirical methods we reviewed in the discussion of analytical models. This work therefore remains an example of a structure where a multiscale approach is needed.

In this specific case, the multiscale approach should provide methods of evaluating electronic, vibrational, and photonic band structures as a function of heterojunction composition and geometry.

*3.2. LIMA: Enhanced light matter interaction*

A second example which we will mention in less detail is the SILVACO modelling in the LIMA European FP7 project [11]. This project has applied plasmonic nanoparticles to Si back-contact solar cells coated with a Si-quantum dot (QD) downshifting layer shown in figure 6. The project was a success in that it demonstrated the first successful integration of Si-QDs fabrication with interdigitated back contact (IBC) solar cells, and the first demonstration of enhanced efficiency from downshifting in such a system.

The plasmonic layer, however, was found to reduce efficiency due to absorption losses in the short wavelength, related to Fano interference near the plasmonic resonance at approximately 450nm. In the final stages of the project, the team did demonstrate enhanced short circuit current by placing the plasmonic particles on the rear surface of the structure, but were unable to resolve the resulting shunting problems before the end of the FP7 project.

The modelling of this structure was a combination of first-principles exact solution of Maxwell's equations [10] developed in the project and semi-empirical modelling of QD downshifting layers [12] to provide the optical response of these layers as inputs to the IBC modelling. The plasmonic response of metallic nano-particles is, in principle, achievable with SILVACO's FDTD capability, but was found limited by issues of computational power, leading the project to rely on the solution settled on, which was the integration of results from third party plasmonic calculations provided by partners in the project. The description of quantum structures in ATLAS being restricted at present to quantum wells, this aspect was similarly developed separately by project partners.

This piece-meal approach [13], while successful in reproducing experimental opto-electronic characteristics of the LIMA cells nevertheless faces limitations of generality. It underlines in particular the incomplete description of quantum confined structures in the numerical packages available.

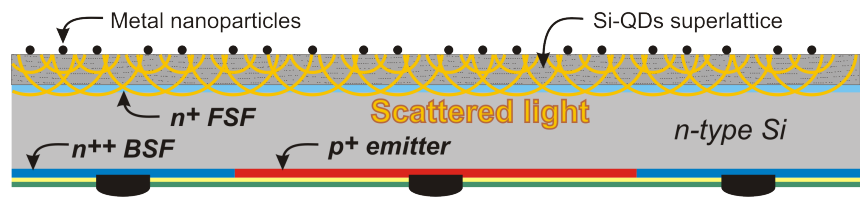

**Figure 6.** The LIMA cell: enhancing Si cell light interaction using plasmonic techniques coupled with quantum dot downshifting

## 4. Multiscale techniques in the MultiscaleSolar project

The previous sections have given some specific case studies illustrating the need for multiscale techniques. Contributors to this work have proposed a European Cooperation in Science and Technology Action along these lines which this paper introduces, and which is running from June 2015 [14].

The primary goals of this Action are the development of both multiscale modelling and multiscale validation of this modelling for optoelectronics, with a focus on photovoltaics which is the origin of the consortium. The objects of study include not only those elements mentioned in this paper, but are open to the identification of any relevant systems where a multiscale analysis is required. These include the full range of so-called third generation photovoltaic structures, and therefore include organic as well as inorganic systems.

The Action is structured in four workgroups [15]. Three of these workgroups focus successively on atomistic, on mesoscale, and on device or macroscales and aim to combine experimental and theoretical developments yielding the multiscale analysis. The action includes an important societal aspect which is a fourth workgroup devoted to industrial perspectives, which monitors the activities of the first three workgroups and evaluates the societal relevance of their progress.

The ultimate aim of this multiscale analysis is the evaluation of third generation concepts such as hot carrier and intermediate band solar cells. The potential of these very promising structures has been described phenomenologically but lacks a quantitative description enabling strategic planning in their application to the world's pressing energy requirements.

## 5. Conclusions

We have seen that semi-empirical analytical models of nanostructures are capable of yielding a refined understanding of the physics of opto-electronic devices. In the QWSC case, an analysis breaking down device characteristics into fundamental recombination processes is able to model characteristics with a single free parameter by simultaneously fitting dark currents and also wavelength dependent light current and characteristics. This shows an advantage of QWSCs which is their radiative dominance in a regime where the same is not true of bulk devices. The limitations of this approach become apparent however when the physical model is stretched to its limits and becomes incapable of explaining why certain phenomena seem to be present. This is the case for apparent hot carrier signals. At this point is it clear that the macroscopic semi-empirical approach must be replaced by a modelling approach capable of correctly accounting for transport, vibrational, and optical modifications of material properties in nanostructured devices. This is a first example of the need for multiscale techniques.

A study of numerical techniques shows a closely related issue. In this case, the more flexible numerical approach in principle allows arbitrary specification of device structure, limited only by knowledge of materials. Leaving aside inessential considerations of computing power, what we find is that available numerical models tend to rely on similar semi-empirical methods to the analytical techniques we have seen. The physical models available are more complete, but we find that materials properties are frequently parameterised on the basis of experimental characterisation. The limitations explored in this paper include the absence of strain parameterisations in one case, and the absence of quantum nanostructure formalisms (with the exception of quantum wells) in the other. On this point it must be emphasised that there are models which are capable of a wider range of physical models which include strain effects and 0D to 2D quantum structures, but which at present fall short on the device simulation framework and are incapable of simulating devices based on such structures.

These examples are illustrations of the thinking that forms the basis of a drive towards multiscale modelling and characterisation under the European COST program within the MultiscaleSolar COST Action. The subjects of this COST Action include organic and inorganic concepts for low cost and high efficiency photovoltaic energy primarily, with an application to optoelectronics generally. The aim of the COST Action is specifically to answer long-standing questions regarding the achievable performance in concepts which have been proposed for some years such as hot carrier cells, intermediate band solar cells, and organic solar cells. The multiscale approach may unlock the potential of these structures.

**Acknowledgments**

The Multispectral Solar cells on Silicon (MULTISOLSI) project is funded by the French Agence Nationale pour la Recherche under the program ANR PROGELEC 2011 ref. ANR-11-PRGE-0009. The COST Action MultiscaleSolar (MP1406) is a programme supported by the Cooperation in Science and Technology Association.